\documentstyle[12pt,prd,aps,epsfig,preprint]{revtex}

\begin{document}
\title{{\bf Polynomial Solution of }${\cal PT}${\bf /Non-}${\cal PT}${\bf %
-Symmetric and Non-Hermitian Generalized Woods-Saxon Potential via
Nikiforov-Uvarov Method}}
\author{Sameer M. Ikhdair\thanks{%
sameer@neu.edu.tr} and \ Ramazan Sever\thanks{%
sever@metu.edu.tr}}
\address{$^{\ast }$Department of Electrical and Electronics Engineering, Near East\\
University, Nicosia, North Cyprus, Mersin 10, Turkey.\\
$^{\dagger }$Department of Physics, Middle East Technical University,\\
Ankara, Turkey.}
\date{\today
}
\maketitle
\pacs{}

\begin{abstract}
Using the Nikiforov-Uvarov method, the bound state energy eigenvalues and
eigenfunctions of the ${\cal PT}$-/non-${\cal PT}$-symmetric and
non-Hermitian generalized Woods-Saxon (WS) potential with the real and
complex-valued energy levels are obtained in terms of the Jacobi
polynomials. According to the ${\cal PT}$-symmetric quantum mechanics, we
exactly solved the time-independent Schr\"{o}dinger equation with same
potential for the ${\rm s}$-states and also for any $l$-state as well. It is
shown that the results are in good agreement with the ones obtained before.

Keywords: Energy Eigenvalues and Eigenfunctions; Generalized Woods-Saxon
Potential; ${\cal PT}$-symmetry, Non-Hermitian potential; Nikiforov-Uvarov
Method.

PACS NUMBER(S): 03.65.-w; 02.30.Gp; 03.65.Ge; 68.49.-h; 24.10.Ht; 03.65.Db;
12.39.Pn; 71.15.Dx; 02.30.Fn
\end{abstract}

\bigskip

\section{Introduction}

\noindent A large variety of potentials with the real or complex form are
encountered in various fields of physics. A consistent physical theory of
quantum mechanics in terms of Hermitian Hamiltonians is built on a complex
Hamiltonian that is not Hermitian, but the energy levels are real and
positive as a consequence of ${\cal PT}{\rm -}$symmetry. By definition, a $%
{\cal PT}{\rm -}$symmetric Hamiltonian $H$ satisfies $[{\cal PT},H]=0,$
i.e., ${\cal PT}H{\cal (PT)}^{-1}={\cal (PT)}^{-1}H{\cal PT}$ $=H,$ where $%
{\cal P}$ and{\rm \ }${\cal T}$ are, respectively, the operators of parity
(or space) and time-reversal transformations. These are defined according to
${\cal P}$ $x{\cal P}$ $=-x,$ ${\cal P}$ $p{\cal P}$ $={\cal T}p{\cal T}$ $%
=-p$ and ${\cal T}iI{\cal T}$ $=-iI$ $\ $where $x,p,$ and $I,$ are,
respectively, the position, momentum and identity operatos acting on the
Hilbert space ${\cal H}$ $=L^{2}({\cal R})$ and $i=\sqrt{-1}.$ Note that
this applies only for the system whose classical position $x$ and momentum $%
p $ are real $[1,2].$ It is also known that ${\cal PT}{\rm -}$symmetry does
not necessarily lead to completely real spectrum, and there are several
potentials where part or all of the energy spectrums are complex. The
Schr\"{o}dinger equation (${\rm SE}$) for the real (Hermitian) potentials
are investigated to generate the real energy eigenvalues which are of much
interest [1]. Bender {\it et al.} for the first time and latter others have
investigated several complex potentials on the ${\cal PT}{\rm -}$symmetric
quantum mechanics. The main reason for the growing recent interest in ${\cal %
PT}{\rm -}$symmetry.[1] is that the eigenvalues (spectrum) of every
Hamiltonian is real (${\cal PT}{\rm -}$symmetry is exact) or come in complex
conjugate pairs of complex eigenvalues (${\cal PT}{\rm -}$symmetry is
spontaneously broken) [1,2,3,4,5,6,7,8,9]. Afterwards, non-Hermitian
Hamiltonians with real or complex spectra have been studied by using
numerical and analytical techniques [10,11,12,13,14].

Various different techniques have been employed in solving the above
mentioned potential cases. One of these methods which makes it possible to
present the theory of special functions by starting from a differential
equation has been developed by Nikiforov and Uvarov (NU) method [15]. This
method is based on solving the time-independent ${\rm SE}$ by reducing it
into a generalized equation of hypergeometric form. Exact solution of SE for
central potentials has generated much interest in recent years. So far,
these potentials are the parabolic type potential [16], the Eckart potential
[17,18], the Fermi-step potential [17,18], the Rosen-Morse potential [19],
the Ginocchio barrier [20], the Scarf barriers [21], the Morse potential
[22] and a potential which interpolates between Morse and Eckart barriers
[23]. Many authors have studied on exponential type potentials
[12,24,25,26,27] and quasi exactly solvable quadratic potentials [28,29,30].
In addition, Dirac, Klein-Gordon, and Duffin-Kemmer-Petiau equations for a
Coulomb type potential are solved by using this method [31,32]. The exact
solutions for these models have been obtained analytically.

So far, we have solved the nonrelativistic and semi-relativistic wave
equations using the statistical model [33], a different approach to the
shifted large 1/N expansion technique [34] and also the shifted large $1/%
\overline{l}$ expansion technique [35] with a wide group of static
phenomenological and QCD-motivated potentials to produce the heavy and light
quarkonium spectra. In addition, the energy eigenvalues of the bound states
of an electron in the general exponential cosine screened Coulomb potential
are obtained using the shifted large 1/N expansion method [36]. In this
work, we solve the ${\rm SE}$ using ${\rm NU}$ method with some well-known $%
{\rm WS}$ potentials [4,13,14,37]. This potential form has been used widely
in analysis of heavy-ion reaction and has enjoyed success [37]. It is
selected for a shell model which can be used for describing metallic
clusters in a successful way and for lighting the central part of the
interaction neutron with one heavy nucleus [38,39].

In the present work, the energy eigenvalues and eigenfunctions of the
Hermitian and non-Hermitian form of the generalized ${\rm WS}$ potential are
calculated by employing the ${\rm NU}$ method.

The contents of this paper is as follows: In Section II we breifly present
the generalized form of Woods-Saxon potential inspired from the SUSYQM. In
Section III, we present Nikiforov-Uvarov method and also the solution of the
Schr\"{o}dinger equation with Hermitian form of the generalized ${\rm WS}$
potential for the $l=0$ and $l\neq 0$ cases. In Section IV, the ${\cal PT}%
{\rm -}$symmetric and non-${\cal PT}{\rm -}$symmetric non-Hermitian
potential forms are also investigated. Shapes of the potential and energy
are studied in Section V. Finally, in Section VI we give results and
conclusions.

\section{Generalized Woods-Saxon Potential}

The motion of the free electrons which have conclusive influence on the
abundance of metallic clusters \ is a vital problem in the nuclear physics.
These electrons are moving in well defined orbitals, around the central
nucleus and in a mean field potential which is produced by the positively
charged ions and the rest of electrons. In the mean field potential, the
details of the potential are described by free parameters such as depth,
width and slope of the potential, which have to be fitted to experimental
observations. The interactions between nuclei are commonly described by
using a potential that consist of the Coulomb and the nuclear potentials.
These potentials are usually taken to be of the ordinary ${\rm WS}$ form
[40] and has been widely used in analysis of heavy-ion reactions [37].
Recently, Arai [41] has introduced the deformed hyperbolic potentials using
the definitions of the deformed hyper functions. As an example, we may
choose the inter-nuclear generalized ${\rm WS}$ of spherically symmetric
form as [42]

\begin{equation}
V(r)=-V_{1}\frac{e^{-\left( \frac{r-R_{0}}{a}\right) }}{1+qe^{-\left( \frac{%
r-R_{0}}{a}\right) }}+V_{2}\frac{e^{-2\left( \frac{r-R_{0}}{a}\right) }}{%
\left( 1+qe^{-\left( \frac{r-R_{0}}{a}\right) }\right) ^{2}},\text{ \ }q\geq
1
\end{equation}
where $r$ stands for the center-of-mass distance between the projectile
nucleus and the target nucleus. The other parameters in the potential $%
R_{0}=r_{0}A^{1/3}$ is the radius of the corresponding spherical nucleus or
the width of the potential, $A$ is the target mass number, $r_{0}$ is the
radius parameter, $V_{1}$ is the potential depth of the Coulombic part
(i.e., the first term on the right hand side of Eq.(1)), $a$ is the surface
diffuseness parameter which is usually adjusted to the experimental values
of ionization energies [37] and finally $V_{2}$ is an introduced parameter
for the second part of Eq.(1) (transforms like potential barrier) [42].
Further, $q$ is the deformation parameter for the strength of the
exponential part other than unity which is arbitrarily taken to be a real
constant within the potential. Its worthwhile to note that the spatial
coordinates in the potential are not deformed and thus the potential still
remains spherical. Therefore, we start with the original ${\rm SE}$:

\begin{equation}
\left( \frac{p^{2}}{2m}+V(r)\right) \psi ({\bf r})=E\psi ({\bf r}),
\end{equation}
where the classical phase space is assumed to be real, i.e., $x$ and $p$ are
the standard Hermitian operators representing the position and momentum of a
particle of mass $m.$ Employing the separation of variables

\begin{equation}
\psi ({\bf r})=\frac{1}{r}R(r)Y(\theta ,\phi ),
\end{equation}
leads to the simple radial ${\rm SE,}$ for all angular momentum states, of
the type

\begin{equation}
-\frac{\hbar ^{2}}{2m}\frac{d^{2}R(r)}{dr^{2}}+\left( V(r)+\frac{l(l+1)\hbar
^{2}}{2mr^{2}}\right) R(r)=ER(r).
\end{equation}
Now, the aim is to solve the last equation for the given potential in Eq.(1)
for the $l=0$ and $l\neq 0$ cases using the Nikiforov-Uvarov method which
will be introduced breifly and employed in section III.

\section{\noindent Applying Nikiforov-Uvarov Method to the Schr\"{o}dinger
Equation}

The Nikiforov-Uvarov (${\rm NU}$) method provides us an exact solution of
Eq.(4) for certain kind of potentials among them the one given in Eq.(1)
[15]. This method is based upon the solutions of general second order linear
differential equation with special orthogonal functions [43]. For a given
real or complex potentials, the ${\cal PT}{\rm -}$symmetric ${\rm SE}$ in
one dimension is reduced to a generalized equation of hypergeometric type
with an appropriate $s=s(x)$ coordinate transformation. Thus, it takes the
form:

\begin{equation}
\psi ^{\prime \prime }(s)+\frac{\widetilde{\tau }(s)}{\sigma (s)}\psi
^{\prime }(s)+\frac{\widetilde{\sigma }(s)}{\sigma ^{2}(s)}\psi (s)=0,
\end{equation}
where $\sigma (s)$ and $\widetilde{\sigma }(s)$ are polynomials, at most of
second-degree, and $\widetilde{\tau }(s)$ is of a first-degree polynomial.
To find a particular solution for Eq. (5) by separation of variables, we use
the transformation given by

\begin{equation}
\psi (s)=\phi (s)y(s).
\end{equation}
This reduces ${\rm SE}$, Eq.(5), into an equation of hypergeometric type:

\begin{equation}
\sigma (s)y^{\prime \prime }(s)+\tau (s)y^{\prime }(s)+\lambda y(s)=0,
\end{equation}
where $\phi (s)$ is found to \i satisfy the condition $\phi ^{\prime
}(s)/\phi (s)=\pi (s)/\sigma (s).$ Further, $y(s)$ is the hypergeometric
type function whose polynomial solutions are given by Rodrigues relation

\begin{equation}
y_{n}(s)=\frac{B_{n}}{\rho (s)}\frac{d^{n}}{ds^{n}}\left[ \sigma ^{n}(s)\rho
(s)\right] ,
\end{equation}
where $B_{n}$ is a normalizing constant and the weight function $\rho (s)$
must satisfy the condition [15]

\begin{equation}
\left( \sigma (s)\rho (s)\right) ^{\prime }=\tau (s)\rho (s).
\end{equation}
The function $\pi (s)$ and the parameter $\lambda $ required for this method
are defined as
\begin{equation}
\pi (s)=\frac{\sigma ^{\prime }(s)-\widetilde{\tau }(s)}{2}\pm \sqrt{\left(
\frac{\sigma ^{\prime }(s)-\widetilde{\tau }(s)}{2}\right) ^{2}-\widetilde{%
\sigma }(s)+k\sigma (s)},
\end{equation}
and

\begin{equation}
\lambda =k+\pi ^{\prime }(s).
\end{equation}
Here, $\pi (s)$ is a polynomial with the parameter $s$ and the determination
of $k$ is the essential point in the calculation of $\pi (s).$ Thus, for the
determination of $k,$ the discriminant under the square root is being set
equal to zero and the resulting second-order polynomial has to be solved for
its roots $k_{+,-}$. Hence, a new eigenvalue equation for the ${\rm SE}$
becomes

\begin{equation}
\lambda _{n}+n\tau ^{\prime }(s)+\frac{n\left( n-1\right) }{2}\sigma
^{\prime \prime }(s)=0,\text{ \ \ \ \ \ \ }\left( n=0,1,2,...\right)
\end{equation}
where

\begin{equation}
\tau (s)=\widetilde{\tau }(s)+2\pi (s),
\end{equation}
and it will have a negative derivative. Therefore, we start solving ${\rm SE}
$ for the ${\cal PT}$ generalized ${\rm WS}$ following the ${\rm NU}$ for $%
l=0$ and $l\neq 0$ cases as follows:

\subsection{Solution for the $l=0$ case}

In order to calculate the energy eigenvalues and the corresponding
eigenfunctions, the Hermitian real-valued potential form given by Eq.(1) is
substituted into the one$-$dimensional ${\cal PT}{\rm -}$symmetrical ${\rm SE%
}$ with the zero angular momentum states,

\begin{equation}
R^{\prime \prime }(x)+\frac{2m}{\hbar ^{2}}\left[ E+\frac{V_{1}e^{-\alpha x}%
}{1+qe^{-\alpha x}}-\frac{V_{2}e^{-2\alpha x}}{\left( 1+qe^{-\alpha
x}\right) ^{2}}\right] R(x)=0,
\end{equation}
where some assignments of one-dimensional parameter $x=r-R_{0}$ and $\alpha
=1/a$ are done.

Now, rewriting Eq.(14) by employing the convenient transformation, $%
s(x)=-e^{-\alpha x},$ then the ${\cal PT}{\rm -}$symmetrical Hermitian
one-dimensional ${\rm SE}$ becomes

\begin{equation}
\frac{d^{2}R(s)}{ds^{2}}+\frac{1}{s}\frac{dR(s)}{ds}+\frac{2m}{\hbar
^{2}\alpha ^{2}s^{2}}\left[ E-\frac{V_{1}s}{1-qs}-\frac{V_{2}s^{2}}{\left(
1-qs\right) ^{2}}\right] R(s)=0,
\end{equation}
and also introducing the given dimensionless parameters:

\begin{equation}
\epsilon =-\frac{2mE}{\hbar ^{2}\alpha ^{2}}>0\text{ \ \ (}E<0),\text{ \ \ }%
\beta =\frac{2mV_{1}}{\hbar ^{2}\alpha ^{2}}\text{ \ \ (}\beta >0),\text{ \ }%
\gamma =\frac{2mV_{2}}{\hbar ^{2}\alpha ^{2}}\text{ \ (}\gamma >0),
\end{equation}
finally it leads into the following simple hypergeometric form given by

\begin{equation}
\frac{d^{2}R(s)}{ds^{2}}+\frac{1-qs}{s(1-qs)}\frac{dR(s)}{ds}+\frac{1}{\left[
s\left( 1-qs\right) \right] ^{2}}\times \left[ \left( -\epsilon q^{2}+\beta
q-\gamma \right) s^{2}+\left( 2\epsilon q-\beta \right) s-\epsilon \right]
R(s)=0.
\end{equation}
Hence, comparing the last equation with the generalized hypergeometric type,
Eq.(5), we obtain the associated polynomials as

\begin{equation}
\widetilde{\tau }(s)=1-qs,\text{ \ \ \ }\sigma (s)=s(1-qs),\text{ \ \ }%
\widetilde{\sigma }(s)=\left( -\epsilon q^{2}+\beta q-\gamma \right)
s^{2}+\left( 2\epsilon q-\beta \right) s-\epsilon .
\end{equation}
When these polynomials are substituted into Eq.(10), with $\sigma ^{\prime
}(s)=1-2qs,$ we obtain

\begin{equation}
\pi (s)=-\frac{qs}{2}\pm \frac{1}{2}\sqrt{\left( q^{2}+4\epsilon
q^{2}-4\beta q+4\gamma -4kq\right) s^{2}+4\left( \beta -2\epsilon q+k\right)
s+4\epsilon }.
\end{equation}
Further, the discriminant of the upper expression under the square root has
to be set equal to zero. Therefore, it becomes

\begin{equation}
\Delta =\left[ 4\left( \beta -2\epsilon q+k\right) \right] ^{2}-4\times
4\epsilon \left( q^{2}+4\epsilon q^{2}-4\beta q+4\gamma -4kq\right) =0.
\end{equation}
Solving Eq.(20) for the constant $k,$ we get the double roots as $%
k_{+,-}=-\beta \pm q\sqrt{\epsilon \left( 1+\frac{4\gamma }{q^{2}}\right) }.$
Thus, substituting these values for each $k$ into Eq.(19), we obtain

\begin{equation}
\pi (s)=-\frac{qs}{2}\pm \frac{1}{2}\left\{
\begin{array}{c}
\left[ \left( 2\sqrt{\epsilon }-\sqrt{1+\frac{4\gamma }{q^{2}}}\right) qs-2%
\sqrt{\epsilon }\right] ;\text{ \ \ \ for \ \ }k_{+}=-\beta +q\sqrt{\epsilon
\left( 1+\frac{4\gamma }{q^{2}}\right) }, \\
\left[ \left( 2\sqrt{\epsilon }+\sqrt{1+\frac{4\gamma }{q^{2}}}\right) qs-2%
\sqrt{\epsilon }\right] ;\text{ \ \ for \ \ }k_{-}=-\beta -q\sqrt{\epsilon
\left( 1+\frac{4\gamma }{q^{2}}\right) }.
\end{array}
\right.
\end{equation}
Hence, making the following choice for the polynomial $\pi (s)$ as

\begin{equation}
\pi (s)=-\frac{qs}{2}-\frac{1}{2}\left[ \left( 2\sqrt{\epsilon }+\sqrt{1+%
\frac{4\gamma }{q^{2}}}\right) qs-2\sqrt{\epsilon }\right] ,
\end{equation}
gives the function:

\begin{equation}
\tau (s)=1-2qs-\left[ \left( 2\sqrt{\epsilon }+\sqrt{1+\frac{4\gamma }{q^{2}}%
}\right) qs-2\sqrt{\epsilon }\right] ,
\end{equation}
which has a \ negative derivative of the form $\tau (s)=-\left( 2+2\sqrt{%
\epsilon }+\sqrt{1+\frac{4\gamma }{q^{2}}}\right) q.$ Thus, from Eq.(11) and
Eq.(12), we find

\begin{equation}
\lambda =-\beta -\frac{q}{2}\left( 1+2\sqrt{\epsilon }\right) \left( 1+\sqrt{%
1+\frac{4\gamma }{q^{2}}}\right) ,
\end{equation}
and

\begin{equation}
\lambda _{n}=\left( 2+2\sqrt{\epsilon }+\sqrt{1+\frac{4\gamma }{q^{2}}}%
\right) nq+n(n-1)q.
\end{equation}
After setting $\lambda _{n}=\lambda $ and solving for $\epsilon ,$ we find:

\begin{equation}
\epsilon _{n}(\alpha ,q)=\left[ \frac{1+2n}{2}-\frac{\left( n(n+1)-\frac{%
\beta }{q}\right) }{1+2n+\sqrt{1+\frac{4\gamma }{q^{2}}}}\right] ^{2}.
\end{equation}
Therefore, substituting the values of $\epsilon ,$ $\beta $ and $\gamma $
into Eq.(26) together with the transformation $\alpha =1/a$, one can
immediately determine the exact energy eigenvalues $E_{n}(a,q)$ as

\begin{equation}
\text{E}_{n}(a,q)=-\frac{\hbar ^{2}}{2ma^{2}}\left[ \frac{1+2n}{2}-\frac{%
\left( n(n+1)-\frac{2ma^{2}V_{1}}{\hbar ^{2}q}\right) }{1+2n+\sqrt{1+\frac{%
8ma^{2}V_{2}}{\hbar ^{2}q^{2}}}}\right] ^{2},\text{ \ \ }0\leq n<\infty ,%
\text{ \ \ }q\geq 1.
\end{equation}
Following Ref.[36], in atomic units ($\hbar =m=c=e=1),$ Eq.(27) turns out to
be

\begin{equation}
\text{E}_{n}(a,q)=-\frac{1}{2a^{2}}\left[ \frac{1+2n}{2}-\frac{\left( n(n+1)-%
\frac{2a^{2}V_{1}}{q}\right) }{1+2n+\sqrt{1+\frac{8a^{2}V_{2}}{q^{2}}}}%
\right] ^{2},\text{ \ \ }0\leq n<\infty ,\text{ \ \ }q\geq 1.
\end{equation}
The above equation indicates that we deal with a family of the generalized $%
{\rm WS}$ spherical potential. Of course it is clear that by imposing
appropriate changes in the parameters $a$ and $V_{1},$ the index $n$
describes the quantization for the bound energy states. In addition, if the
parameter $V_{2}$ in Eq.(28) is adjusted to zero (for ${\rm s-}$state),
solution reduces to the form obtained for the standard ${\rm WS}$ potential
without regarding of the centrifugal barrier potential and the deformation
parameter $q$ in Eq.(4)$.$\footnote{%
The exact bound energy eigenvalues for the ${\rm s-}$states, in atomic
units, is E$_{n}(a=1,q\rightarrow 1)=-\frac{1}{8}\left[ n+1+\frac{2V_{1}}{%
(n+1)}\right] ^{2},$ \ \ which is in agreement with previous works on SUSYQM
(cf. formula (40) in Ref.[42]).}

Let us now find the corresponding wavefunctions. Applying the ${\rm NU}$
method, the polynomial solutions of the hypergeometric function $y(s)$
depends on the determination of weight function $\rho (s)$ which is found to
be

\begin{equation}
\rho (s)=(1-qs)^{\eta -1}s^{2\sqrt{\epsilon }};\text{ \ \ \ \ }\eta =1+\sqrt{%
1+\frac{4\gamma }{q^{2}}}.
\end{equation}
Substituting into the Rodrigues relation given in Eq.(8), the eigenfunctions
are obtained in the following form

\begin{equation}
y_{n,q}(s)=C_{n}(1-qs)^{-(\eta -1)}s^{-2\sqrt{\epsilon }}\frac{d^{n}}{ds^{n}}%
\left[ \left( 1-qs\right) ^{n+\eta -1}s^{n+2\sqrt{\epsilon }}\right] ,
\end{equation}
where $C_{n}$ stands for the normalization constant and its value is $1/n!.$
In the limit $q\rightarrow 1,$ the polynomial solutions of $\ y_{n}(s)$ are
expressed in terms of Jacobi Polynomials, which is one of the classical
orthogonal polynomials, with weight function $(1-qs)^{\eta -1}s^{2\sqrt{%
\epsilon }}$ in the closed interval $\left[ 0,1\right] ,$ yielding $%
y_{n,1}(s)\simeq P_{n}^{(2\sqrt{\epsilon },\eta -1)}(1-2s)$ [43]. Finally,
the other part of the wave function in Eq.(6) (with $q\rightarrow 1$) is
found to be

\begin{equation}
\phi (s)=(1-s)^{\mu }s^{\sqrt{\epsilon }},\text{ \ \ }\mu =\eta /2.
\end{equation}
Combining the Jacobi polynomials and $\phi (s)$ in Eq.(31), the ${\rm s-}$%
wave functions ($l=0)$ could be determined as

\begin{equation}
R_{n}(s)=D_{n}s^{\sqrt{\epsilon }}(1-s)^{\eta /2}P_{n}^{(2\sqrt{\epsilon }%
,\eta -1)}(1-2s),
\end{equation}
with $s=-e^{-\alpha x}$ and $D_{n}$ is a new normalization constant.

\subsection{Solution for the $l\neq 0$ case}

The Hamiltonian for the generalized ${\rm WS}$ potential for the $l\neq 0$
case is

\begin{equation}
H=\frac{p^{2}}{2m}-V_{1}\frac{e^{-\left( \frac{r-R_{0}}{a}\right) }}{%
1+qe^{-\left( \frac{r-R_{0}}{a}\right) }}+\frac{l(l+1)\hbar ^{2}}{2mr^{2}}.
\end{equation}
To evaluate the spectra of energy eigenvalues and eigenfunctions we
introduce a new effective potential whose form is given as:

\begin{equation}
V_{eff}=-V_{1}\frac{e^{-\left( \frac{r-R_{0}}{a}\right) }}{1+qe^{-\left(
\frac{r-R_{0}}{a}\right) }}+\frac{l(l+1)\hbar ^{2}\alpha ^{2}}{2m}\frac{%
e^{-2\left( \frac{r-R_{0}}{a}\right) }}{\left( 1+qe^{-\left( \frac{r-R_{0}}{a%
}\right) }\right) ^{2}},
\end{equation}
where the second term of the last expression looks like the potential
barrier term of Eq.(33). Thus, comparing Eq.(34) with its counterpart
Eq.(1), one can arrive into the transformation of $V_{2}\rightarrow $ $\frac{%
l(l+1)\hbar ^{2}\alpha ^{2}}{2m}.$ Moreover, Eq.(34) can also be rewriten as

\begin{equation}
V_{eff}=-V_{1}\frac{1}{e^{\left( \frac{r-R_{0}}{a}\right) }+q}+\frac{%
l(l+1)\hbar ^{2}\alpha ^{2}}{2m\left( e^{\left( \frac{r-R_{0}}{a}\right)
}+q\right) ^{2}},
\end{equation}
where $\alpha =(q+1)/R_{0}$ and $\alpha =1/a.$ The lowest energy levels of
the potential in Eq.(34) are now given

\begin{equation}
\text{E}_{n,l}(a,q)=-\frac{\hbar ^{2}}{2ma^{2}}\left[ \frac{1+2n}{2}-\frac{%
\left( n(n+1)-\frac{2ma^{2}V_{1}}{\hbar ^{2}q}\right) }{1+2n+\sqrt{1+\frac{%
4l(l+1)}{q^{2}}}}\right] ^{2},\text{ \ \ }0\leq n<\infty ,\text{ \ \ }q\geq
1.
\end{equation}
For scattering processes, it has been well accepted that the surface
diffuseness parameter $a$ is around $0.63$ $fm$ [37,44,45]. Much larger
diffuseness parameter, ranging between $0.8$ and $1.4$ $fm$ is needed in
order to fit the data [37,46].

\section{Non-Hermitian Potential Forms}

\subsection{${\cal PT}$-symmetric and non-Hermitian generalized WS case}

For ${\cal PT}{\rm -}$symmetric and non-Hermitian potential case, we take
the potential parameters, in Eq.(1), $V_{1},$ $V_{2},$ and $q$ as real and $%
\alpha \rightarrow i\alpha _{I}$ is purely imaginary$.$ In this case, the
potential turns out to be
\begin{equation}
V(x)=-V_{1}\left( \frac{q+\cos \alpha _{I}x-i\sin \alpha _{I}x}{%
1+q^{2}+2q\cos \alpha _{I}x}\right) +V_{2}\left( \frac{q+\cos \alpha
_{I}x-i\sin \alpha _{I}x}{1+q^{2}+2q\cos \alpha _{I}x}\right) ^{2}.
\end{equation}
By substituting this potential into Eq.(14) and making similar operations in
obtaining Eq.(28), one can easily get the energy eigenvalues, in atomic
units, as

\begin{equation}
E_{n}(\alpha _{I},q)=\frac{1}{2}\left[ \frac{\alpha _{I}(1+2n)}{2}-\frac{%
\left( n(n+1)\alpha _{I}+\frac{2V_{1}}{\alpha _{I}q}\right) }{1+2n+\sqrt{1-%
\frac{8V_{2}}{\alpha _{I}^{2}q^{2}}}}\right] ^{2},
\end{equation}
and

\begin{equation}
\lambda =\beta -\frac{q}{2}\left( 1+i2\sqrt{\epsilon }\right) \left( 1+\sqrt{%
1-\frac{8mV_{2}}{\hbar ^{2}\alpha _{I}^{2}q^{2}}}\right) ,
\end{equation}

\begin{equation}
\lambda _{n}=2\left( 1+i\sqrt{\epsilon }+\frac{1}{2}\sqrt{1-\frac{8mV_{2}}{%
\hbar ^{2}\alpha _{I}^{2}q^{2}}}\right) nq+n(n-1)q.
\end{equation}
Here $\alpha _{I}$ is an arbitrary real parameter and $i=\sqrt{-1}.$ Thus,
by choosing the parameter $\alpha $ as purely imaginary, we find the energy
eigenvalues obtained for ${\cal PT}{\rm -}$symmetric and non-hermitian
generalized ${\rm WS}$ potential are not similar to Eq.(28). A positive
energy spectra is obtained if and only if \ $n<\sqrt{\frac{2m}{\hbar
^{2}\alpha _{I}^{2}}\left( \frac{V_{1}}{q}-\frac{V_{2}}{q^{2}}\right) }-%
\frac{1}{2}\left( 1+\sqrt{1-\frac{8mV_{2}}{\hbar ^{2}\alpha _{I}^{2}q^{2}}}%
\right) ,$ since the energy eigenvalues of generalized ${\rm WS}$ potential
are negative.\footnote{%
Once we set $V_{2}=0,$ then $n<\sqrt{\frac{2mV_{1}}{\hbar ^{2}\alpha
_{I}^{2}q}}-1$ which agrees with Ref.[13,14].} It appears that the
eigenvalues are always positive real for $V_{2}=0,$ that is, $E_{n}(\alpha
_{I},q)=\frac{1}{2}\left[ \frac{1+n}{2}\alpha _{I}-\frac{V_{1}}{(1+n)\alpha
_{I}q}\right] ^{2},$ but can be complex for $V_{2}>\frac{\alpha _{I}^{2}q^{2}%
}{8}$.\ Since these one--dimensional non-Hermitian Hamiltonian were
invariant under ${\cal PT}{\rm -}$transformation, they possessed real
spectra. Thus, their real spectral properties were linked with their ${\cal %
PT}{\rm -}$symmetry. \ \ \ \ \ \ \ \ \ \ \ \ \ \ \ \ \ \ \ \ \ \

On the other hand, following the procedures in Section III, we obtain an
eigenfunction for the non-Hermitian potential, Eq.(37), as:

\begin{equation}
R_{n}(s)=D_{n}s^{\widetilde{E}_{n}}(1-s)^{\frac{1}{2}\left( 1+\sqrt{%
1-8V_{2}/\alpha _{I}^{2}}\right) }P_{n}^{(2\widetilde{E}_{n},\sqrt{%
1-8V_{2}/\alpha _{I}^{2}})}(1-2s),
\end{equation}
with $\widetilde{E}_{n}=\left[ \frac{1+2n}{2}-\frac{\left( n(n+1)+\frac{%
2V_{1}}{\alpha _{I}^{2}}\right) }{1+2n+\sqrt{1-\frac{8V_{2}}{\alpha _{I}^{2}}%
}}\right] $ and $s=-e^{-i\alpha _{I}x}.$ Its clear that the eigenfunctions
are always complex for all values of $V_{2}.$ Therefore, if $V_{2}=0$ (with $%
q\rightarrow 1),$ we have a simple wavefunction taking the form:

\begin{equation}
R_{n}(x)=D_{n}(-1)^{\widetilde{E}_{n}}e^{-i\alpha _{I}\widetilde{E}%
_{n}x}(1+e^{-i\alpha _{I}x})P_{n}^{(2\widetilde{E}_{n},1)}(1+2e^{-i\alpha
_{I}x}),
\end{equation}
with $\widetilde{E}_{n}=\left( \frac{1+n}{2}-\frac{V_{1}}{(1+n)\alpha
_{I}^{2}}\right) .$

\ \ \ \ \ \ \ \ \ \ \ \ \ \ \ \ \ \ \ \ \ \ \ \ \ \ \ \ \ \ \ \ \ \ \ \ \ \
\ \ \ \ \ \ \ \ \ \ \ \ \ \ \ \ \ \ \ \ \ \ \ \ \ \ \ \ \ \ \ \ \ \ \ \ \ \
\ \ \ \ \ \ \ \ \ \ \ \ \ \ \ \ \ \ \ \ \ \ \ \ \ \ \ \ \ \ \ \ \ \ \

\subsection{Non-${\cal PT}$-symmetric and non-Hermitian generalized WS case\
\ \ \ \ \ \ \ \ \ \ \ \ \ \ \ \ \ \ \ \ \ \ \ \ \ \ \ \ \ \ \ \ \ \ \ \ \ \
\ \ \ \ \ \ \ \ \ \ \ \ \ \ \ \ \ \ \ \ \ \ \ \ }

Another form of the potential is obtained by making the selection of the
potential parameter as pure imaginary, that is, $V_{1}$ is replaced by $%
iV_{1I},$ and $\alpha $ is replaced by $i\alpha _{I}$ but $V_{2}$ remains a
pure real. In this case, the Hamiltonian is non-Hermitian and non-${\cal PT}%
{\rm -}$symmetric having real spectra. The potential turns out to be
\begin{equation}
V(x)=-V_{1I}\left( \frac{\sin \alpha _{I}x+i(q+\cos \alpha _{I}x)}{%
1+q^{2}+2q\cos \alpha _{I}x}\right) +V_{2}\left( \frac{q+\cos \alpha
_{I}x-i\sin \alpha _{I}x}{1+q^{2}+2q\cos \alpha _{I}x}\right) ^{2}.
\end{equation}
By substituting this potential into Eq.(14) and making similar operations in
obtaining Eq.(28), one can easily get the energy eigenvalues, in atomic
units, as

\begin{equation}
E_{n}(\alpha _{I},q)=\frac{1}{2}\left[ \frac{\alpha _{I}(1+2n)}{2}-\frac{%
\left( i\frac{2V_{1I}}{\alpha _{I}q}+n(n+1)\alpha _{I}\right) }{1+2n+\sqrt{1-%
\frac{8V_{2}}{\alpha _{I}^{2}q^{2}}}}\right] ^{2},
\end{equation}
and

\begin{equation}
\lambda =i\beta -\frac{q}{2}\left( 1+i2\sqrt{\epsilon }\right) \left( 1+%
\sqrt{1-\frac{8mV_{2}}{\hbar ^{2}\alpha _{I}^{2}q^{2}}}\right) ,
\end{equation}

\begin{equation}
\lambda _{n}=2\left( 1+i\sqrt{\epsilon }+\frac{1}{2}\sqrt{1-\frac{8mV_{2}}{%
\hbar ^{2}\alpha _{I}^{2}q^{2}}}\right) nq+n(n-1)q.
\end{equation}
The last case has real plus imaginary energy spectra. When we consider the
real part of energy eigenvalues an acceptable result is obtained when $n<%
\sqrt{\frac{2m}{\hbar ^{2}\alpha _{I}^{2}}\left( \frac{iV_{1I}}{q}-\frac{%
V_{2}}{q^{2}}\right) }-\frac{1}{2}\left( 1+\sqrt{1-\frac{8mV_{2}}{\hbar
^{2}\alpha _{I}^{2}q^{2}}}\right) $ condition. However, the energy spectrum
is not seen at the imaginary part of energy eigenvalues, since it is
independent of $n.$\footnote{%
Once we set $V_{2}=0,$ then $n<\sqrt{\frac{i2mV_{1}}{\hbar ^{2}\alpha
_{I}^{2}q}}-1$ which agrees with Ref.[13,14].} \ \ \ \ \ \ \ \ \ \ \ \ \ \ \
\ \ \ \ \ \ \ \ \ \ \ \ \ \ \ \ \ \ \ \ \ \ \ \ \ \ \ \ \ \ \ \ \ \ \ \ \ \
\ \ \ \ \ \ \ \ \ \ \ \ \ \ \ \ \ \ \ \ \ \ \ \ \ \ \ \ \ \ \ \ \ \ \ \ \ \
\ \ \ \ \ \ \ \ \ \ \ \ \ \ \ \ \ \ \ \ \ \ \ \ \ \ \ \ \ \ \ \ \ \ \ \ \ \
\ \ \ \ \ \ \

\section{Potential and Energy Shapes}

It is illustrated in Figure 1, the shape of the generalized ${\rm WS}$
potential given by Eq.(1) for various $q=1,3$ and $7$ values. This is done
by fixing the potential parameters $V_{1}=50$ $MeV,$ $V_{2}=10$ $MeV,$ $%
r_{0}=1.285$ $fm,$ $A=56$ and $a=0.65$ $fm$ [37,44,45]$.$ The empirical
values found by Perey et al. [47] $r_{0}=1.285$ $fm,$ $a=0.65$ $fm$ and $A=56
$ which is the geometric average of the target nucleus mass number $44\leq
A\leq 72$ [48] are used to draw the potential in Figure 2 for various values
of $V_{2}=10,$ 50$,$ $100$ $MeV.$ Further, Figure 3 shows the energy
eigenvalues in atomic units as a function of the discrete level $n$ for
various $a=0.65,$ $0.85,$ 1.05 $fm$ values$.$ Of course, it is clear that by
imposing appropriate changes in the parameter $a,$ the index $n$ describes
the quantization of the bound states and energy spectrum. Some of the
initial energy levels for undeformed case ($q=1)$ value, are presented by
choosing $V_{1}=50$ $MeV,$ $V_{2}=10$ $MeV$ in Figure 4. Its worthwhile to
note that Figure 3 and Figure 4 have no lower bound on the spectrum for both
$V_{2}=0$ and $V_{2}\neq 0$ cases, respectively. In this regard, it is found
that for $V_{2}=0$ case, the energy levels go into lower bound for $0<n<6$
states and into higher bound for $n>6$ states. However, for $V_{2}\neq 0$
case, the energy levels go into lower bound for $0<n<2$ states and higher
bound for $n>2$ states.

\section{Results and Conclusions}

In this article we used ${\rm NU}$ method and solved the radial ${\rm SE}$
for the generalized ${\rm WS}$ potential with the angular momentum $l=0$ $%
(V_{2}=0)$ and $l\neq 0$ $(V_{2}\neq 0).$ A particularly interesting result
of our investigation is that all the ${\cal PT}{\rm -}$symmetric
Hamiltonians with potential parameters remain all purely real have a real
bound energies $E_{n}$ with $n\geq 0$ for Hermitian case and real positive
in contrary to expectation if one lets $\alpha \rightarrow i\alpha _{I}$ in
the generalized ${\rm WS}$ potential. Therefore, for non-Hermitian case, the
spectrum is real for $V_{2}=0$ but complex conjugate for some values of $%
V_{2}\neq 0$ in the generalized ${\rm WS}$ potential$.$ Further, when $%
\alpha $ and $V_{1}$ parameters are purely complex, it is seen that the
number of discrete levels for bound states is given only by the real part of
energy eigenvalues. Thus, for a ${\cal PT}{\rm -}$symmetric Hamiltonians the
exactness of ${\cal PT}{\rm -}$symmetry implies the reality of spectrum.
More specifically, if an eigenfunction $R_{n,l}(s)$ is a ${\cal PT}{\rm -}$%
invariant, ${\cal PT}$ $R_{n,l}(s)=R_{n,l}(s),$ then the corresponding
eigenvalue of $E_{n,l}$ is real. The exact ${\cal PT}{\rm -}$symmetry is a
sufficient condition. But for a given ${\cal PT}{\rm -}$symmetric
Hamiltonian, it is not easy to determine the exactness of ${\cal PT}{\rm -}$%
symmetry without actually solving the corresponding radial ${\rm SE}$. In
this regard, the wave functions are physical and energy eigenvalues are in
good agreement with the results obtained by the other methods [42].

On the other hand, the effect of the centrifugal barrier potential which
goes as $1/r^{2}$ was replaced by a term having exactly the original ${\rm WS%
}$ form but of a second degree in order to reproduce its effect. This $l\neq
0$ term has its physical basis arising from the superpotential partner of
the ${\rm WS}$ potential in SUSYQM [42]. Hence, this new barrier term
retakes the exact form of the original potential but with a small perturbed
strength factor; say $V_{2}$ [42,49]$.$ According to the complex quantum
mechanics [50], the eigenvalues of the conversion $\alpha \rightarrow
i\alpha _{I}$ are not simultanously eigenstates of ${\cal PT}{\rm -}$%
operator.

Finally, we point out that the exact results obtained for the generalized $%
{\rm WS}$ potential may have some interesting applications in the study of
different quantum mechanical systems and nuclear scattering.

\acknowledgments This research was partially supported by the
Scientific and Technical Research Council of Turkey. We like to
thank Ibrahim AbuAwwad for his assistance in drawing the Figures. S.
M. Ikhdair acknowledges his wife, Oyoun, and also his son, Musbah,
for their love, encouragement and assistance. Their encouragement
provided the necessary motivation to complete this work.

\newpage

\begin{figure}[tbp]
\caption{Variation of the generalized WS potential as a function
$r.$ The curves are shown for various values of the deformation
parameter $q$} \label{Figure1}
\end{figure}

\bigskip

\begin{figure}[tbp]
\caption{Variation of the generalized WS potential as a function
$r.$ The curves are shown for various values of the perturbed
parameter $V_{2}$.} \label{Figure 2}
\end{figure}


\begin{figure}[tbp]
\caption{The variation of the energy eigenvalues with respect to
the quantum number $n$ with $V_{1}=50$ $MeV$ \ and $V_{2}=0.$ The
curves shown are for various values of the surface diffuseness
parameter $a.$} \label{Figure 3}
\end{figure}


\bigskip

\begin{figure}[tbp]
\caption{The variation of the energy eigenvalues with respect to
the quantum number $n$ with $V_{1}=50$ $MeV$ \ and $V_{2}=10$
$MeV.$ The curves shown are for various values of the surface
diffuseness parameter $a.$} \label{Figure 4}
\end{figure}

\bigskip \baselineskip= 2\baselineskip

\bigskip

\bigskip

\newpage

\begin{figure}
\epsfig{file=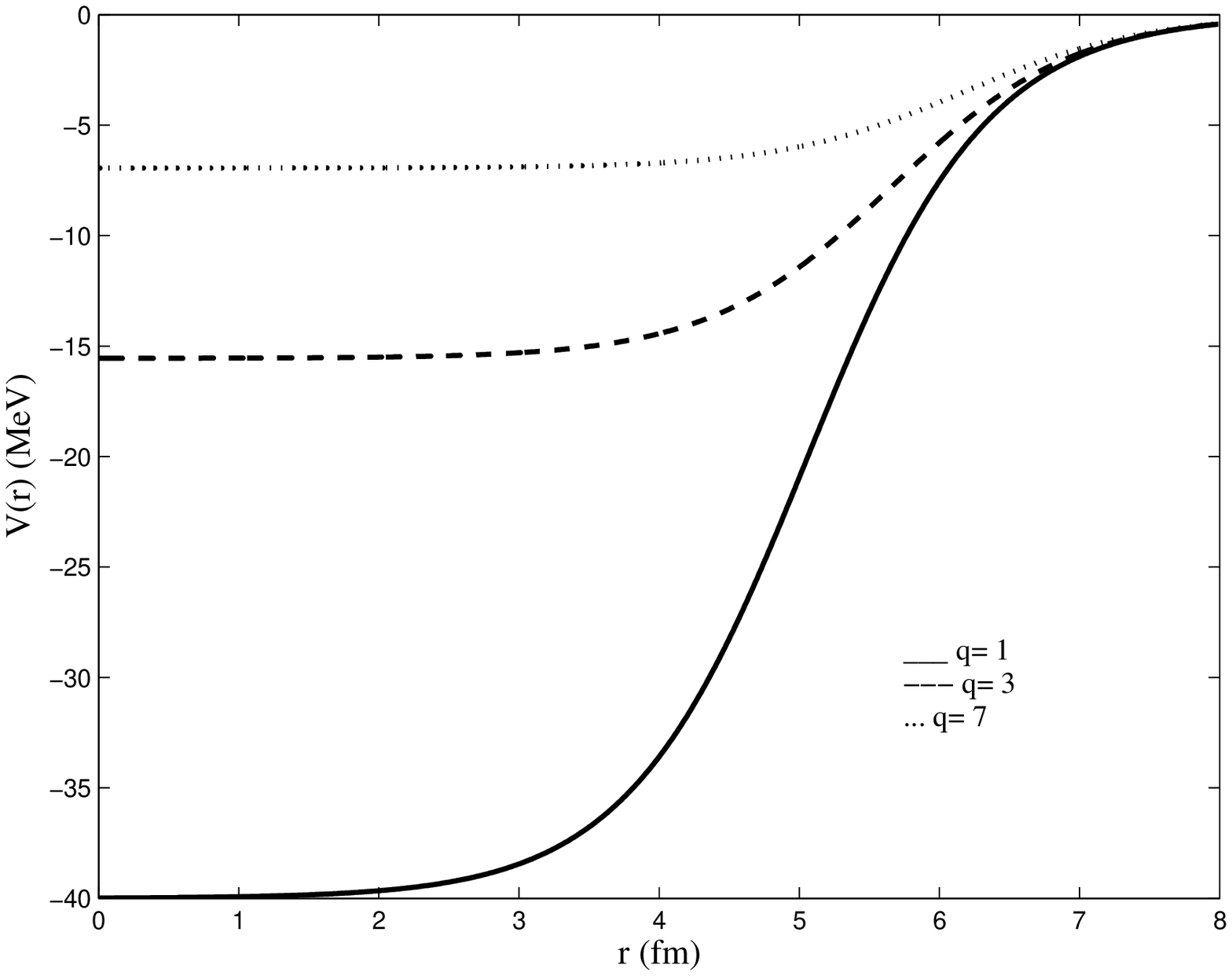,width=17cm,angle=0} \label{Fig1}
\end{figure}

\newpage

\begin{figure}
\epsfig{file=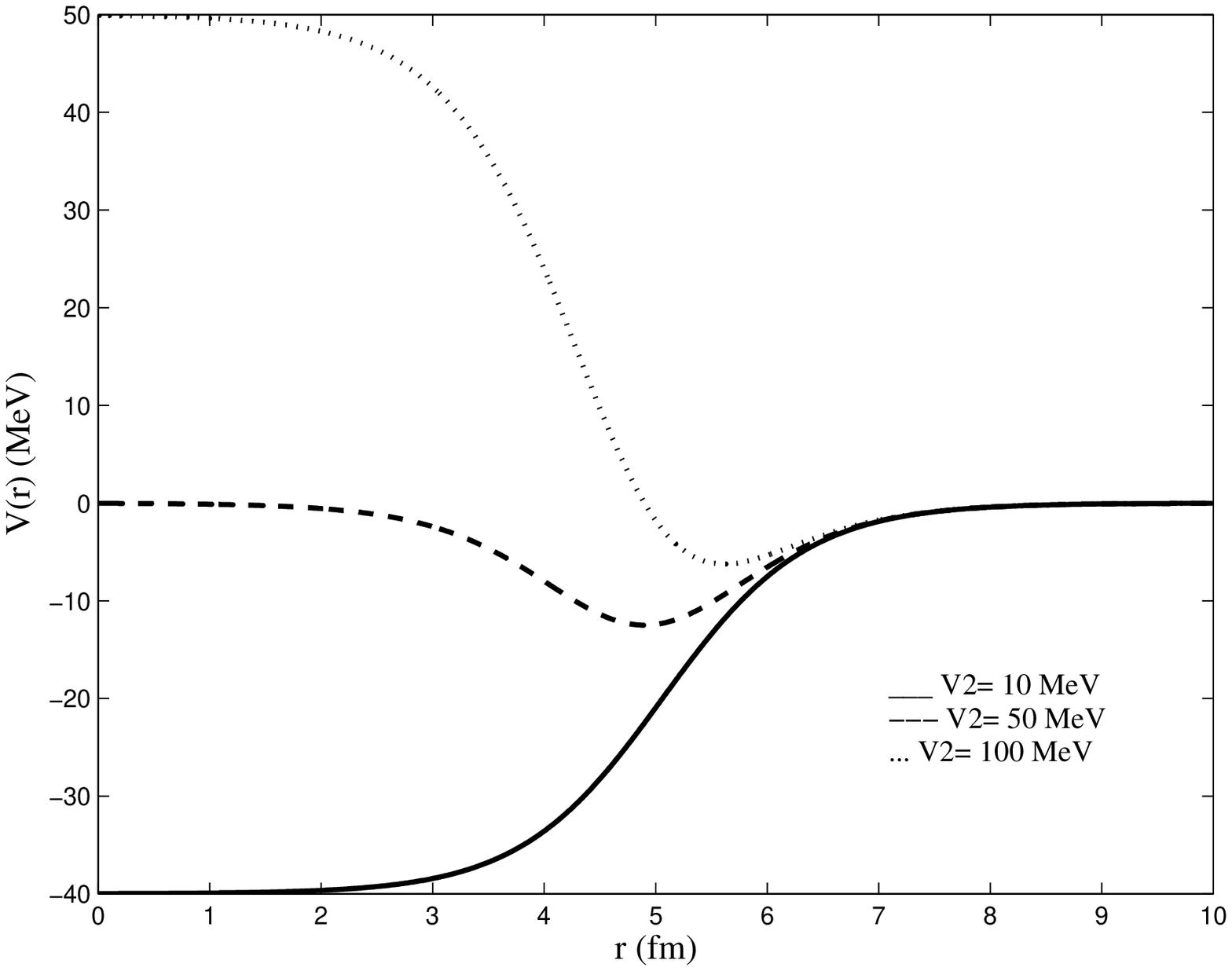,width=17cm,angle=0} \label{Fig2}
\end{figure}

\newpage

\begin{figure}
\epsfig{file=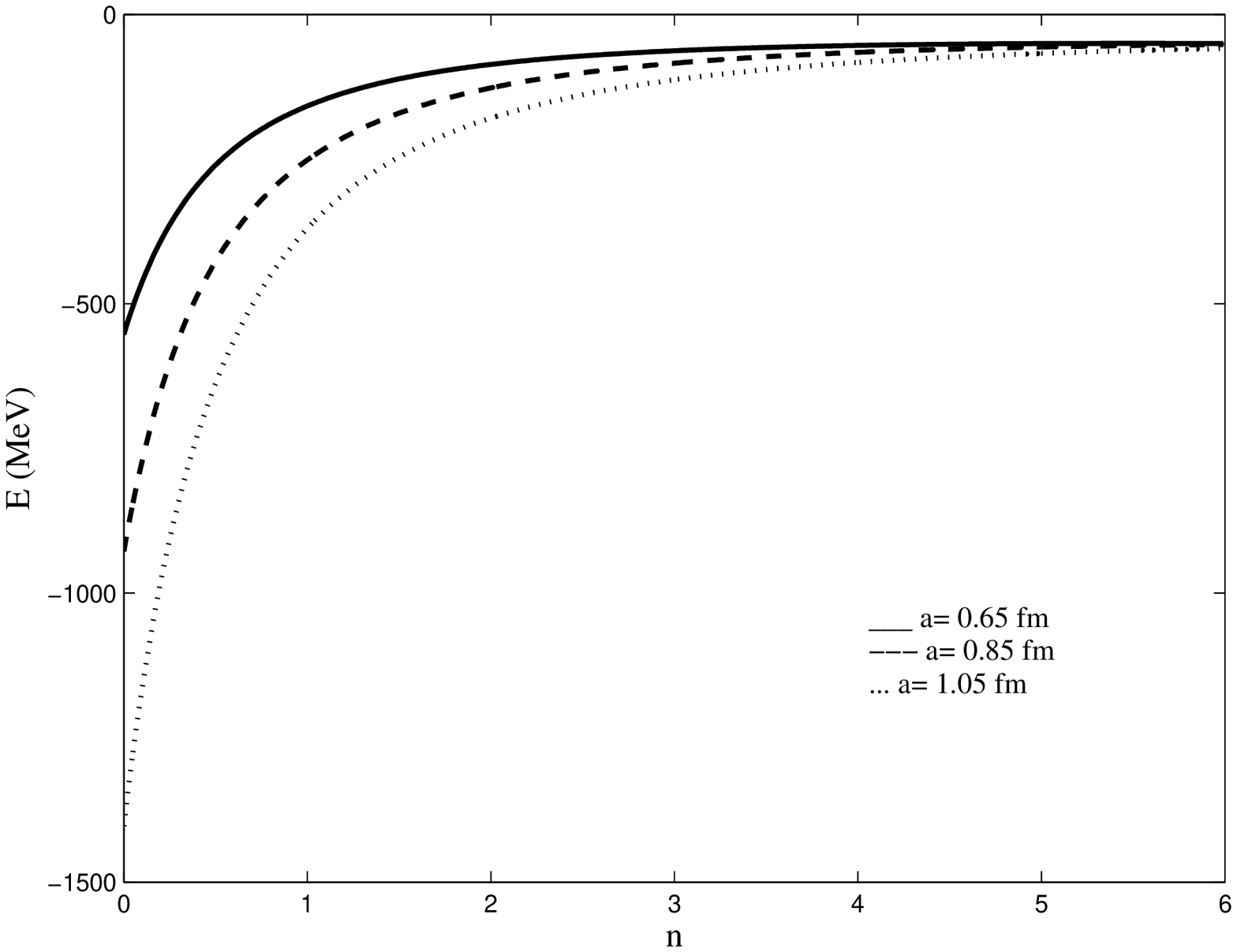,width=17cm,angle=0} \label{Fig3}
\end{figure}

\newpage

\begin{figure}
\epsfig{file=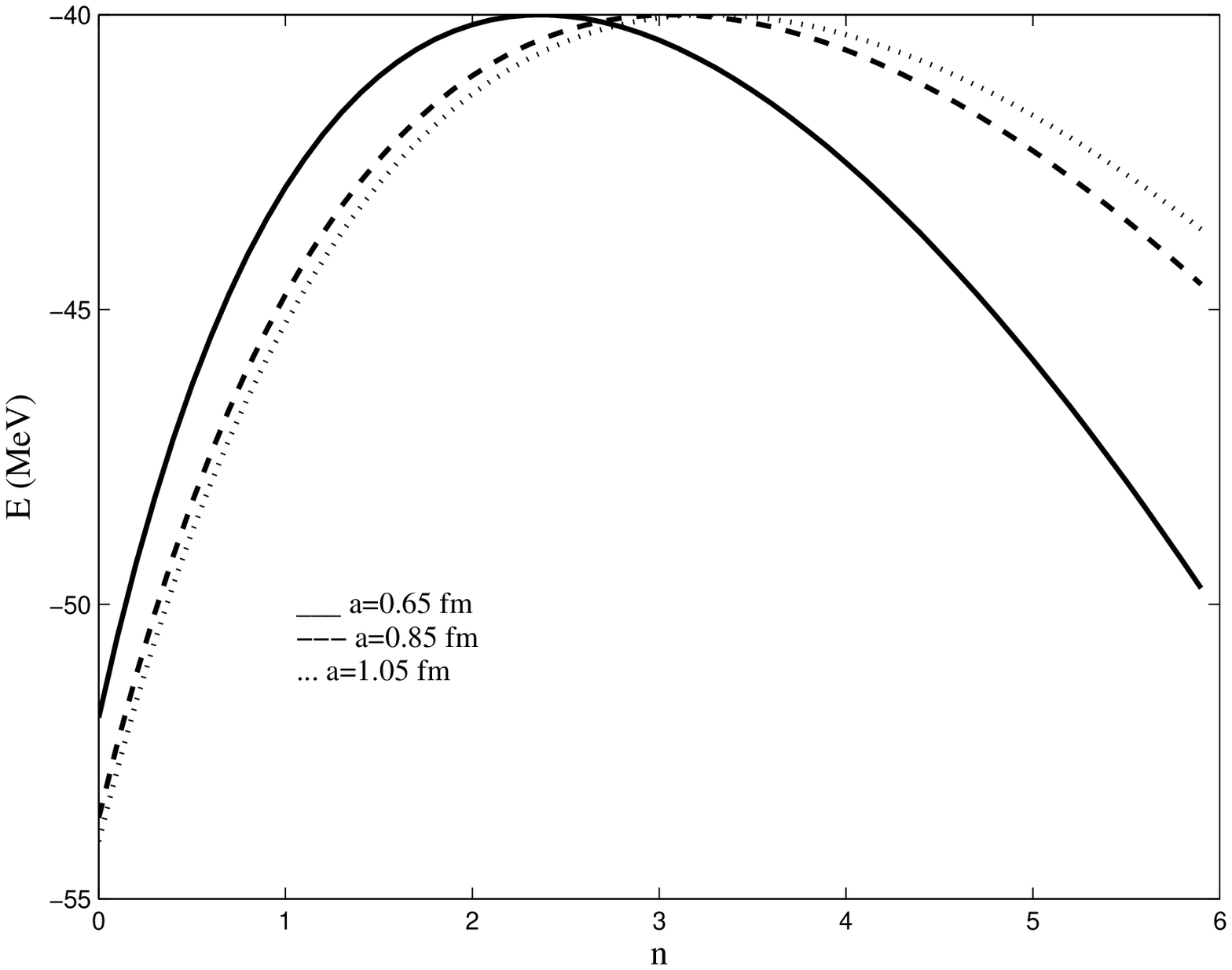,width=17cm,angle=0} \label{Fig4}
\end{figure}
\end{document}